\documentclass[preprint2]{aastex}

\slugcomment{final draft written by KARINO, 3rd. Jan. 2003}

\shorttitle{Dynamical Instability of Differentially Rotating Polytropes}
\shortauthors{Karino and Eriguchi}

\begin{document}

\title 
{Linear Stability Analysis of Differentially Rotating Polytropes \\
---  New results for the m = 2 $f$-mode dynamical instability --- }
\author{Shigeyuki Karino \footnote{\tt karino@provence.c.u-tokyo.ac.jp}
and Yoshiharu Eriguchi}
\affil{Department of Earth Science and Astronomy,
Graduate School of Arts and Sciences, \\
University of Tokyo, Komaba, Meguro, Tokyo 153-8902, Japan}


\begin{abstract}

We have studied the $f$-mode oscillations of differentially rotating 
polytropes by making use of the linear stability analysis. We found that the 
critical values of $T/|W|$ where the dynamical instability against the $m = 2$ 
$f$-mode oscillations sets in decrease down to $T/|W| \sim 0.20$ as the degree 
of differential rotation becomes higher. Here $m$ is an azimuthal mode 
number and $T$ and $W$ are the rotational energy and the gravitational 
potential energy, respectively. This tendency is almost independent of the 
compressibility of the polytropes. These are the {\it first  exact results} 
of the linear stability analysis for the occurrence of the dynamical 
instability against the $m = 2$ $f$-modes.

\end{abstract}


\section{Introduction}

It is well known that rapidly rotating stars suffer from dynamical instability
against nonaxisymmetric oscillations. In particular, the growth rate
for the $m = 2$ or bar-type $f$-mode instability has been considered to be 
largest.  The classical results about this instability were first studied
by Riemann (1860).  Riemann found that uniformly rotating incompressible 
fluids become dynamically unstable when the ratio of the rotational energy, 
$T$, to the absolute value of the gravitational potential energy, $W$, 
exceeds 0.27, i.e. $\beta \equiv T/|W| \ge 0.27$ (see also Bryan 1889).

Around 1970, Ostriker and coworkers developed a new numerical method to solve 
equilibrium configurations of rapidly rotating compressible stars (Ostriker 
\& Mark 1968; Ostriker \& Bodenheimer 1968; Mark 1968; Jackson 1970;
Bodenheimer \& Ostriker 1970, 1973; Bodenheimer 1971). Ostriker and coworkers 
also developed a new technique to explore stabilities of rotating equilibrium 
configurations by extending the tensor-virial technique of Chandrasekhar (
Chandrasekhar 1969; Tassoul \& Ostriker 1968; Ostriker \& Tassoul 1969; 
Ostriker \& Bodenheimer 1973).  Their main result about the $m = 2$ dynamical 
instabilities of compressible stars was that rapidly rotating configurations 
become unstable when the condition $\beta \ge 0.27$ is satisfied irrespective 
of the compressibilities and the rotation laws as far as they 
investigated. Although there was no proof of this "universal" nature of the
critical value $\beta_{\rm critical} \sim 0.27$, most people have believed 
that the critical value for the $m = 2$ $f$-mode dynamical instability is 
0.27. This "belief" has lasted rather long, although there appeared a paper 
that shows the tensor virial technique for stability analysis for 
differentially rotating stars gives neither the necessary condition nor the 
sufficient condition for the stability (Bardeen et al. 1977).

However, the situation has been changing because there appear many recent 
results of hydrodynamical simulations which show that configurations with lower
values of $\beta$ suffer from dynamical instabilities (see e.g. Tohline \& 
Hachisu 1990; Pickett et al. 1996; Toman et al. 1998; Brown 2000; Centrella et 
al. 2001; for simulations of linearized equations see e.g. Liu 2002). 
Rough values of the critical configurations against 
nonaxisymmetric perturbations are $\beta_{\rm critical} \sim 0.16 \ \ 
( m = 2 )$ for self-gravitating toroidal polytropes with the polytropic index 
$N = 1.5$ (Tohline \& Hachisu 1990), $\beta_{\rm critical} \sim 0.2 \ \ 
( m = 2 )$ for $N = 1.5$ polytropic star-disk like configurations (Pickett 
et al. 1996), $\beta_{\rm critical} \sim 0.24 \sim 0.28 \ \ ( m = 2 )$ for 
$N = 5/4, 6/4, 10/4$ polytropes with the rotation laws that the distribution 
of the specific angular momentum is that of uniformly rotating polytropes with 
the same polytropic indices (Toman et al.1998),  $\beta_{\rm critical} 
\sim 0.24 \ \ ( m = 2 )$ for $N = 1.5$ polytropes with the angular velocity of 
the exponential decay with the distance from the rotation axis (Brown 2000), 
and $\beta_{\rm critical} \sim 0.14 \ \ ( m = 1 )$ for toroidal-like 
$N = 3.33$ polytropes (Centrella et al. 2001). Liu found $\beta_{\rm critical} 
\sim 0.25 \ \ ( m = 2 )$ for the angular momentum distribution which could be 
realized after accretion induced collapse of uniformly rotating massive white 
dwarfs, although his simulations were carried out by using  the linearized 
equations.

These results seem to imply that the critical value of $\beta$ for the 
occurrence of the $m = 2$ bar-type dynamical instability cannot be considered 
to be universal any more. In order to see the unstable nature of rotating
stars, we need to investigate sequences of rotating configurations
systematically. For that purpose, the best method would be the linear 
stability analysis rather than hydrodynamical simulations.

Except for non-self-gravitating disks, linear stability analyses for
rapidly rotating self-gravitating stars have been done only for
limited kinds of compressibilities and rotation laws, and for limited types 
of oscillations. Luyten (1990, 1991) found a new dynamical instability 
for differentially rotating polytropes. This instability seems to belong to
the shear instability which is also found in the theory of accretion disks
(see e.g. Papaloizou \& Pringle 1984; Narayan \& Goodman 1989). 
This instability appears for differentially and very slowly rotating stars 
and disappears for differentially but very rapidly rotating stars. However, 
Luyten's analysis was applied only to very limited kinds of rotation laws and 
the displacements were restricted only in the plane parallel to the equator. 
Very recently, Shibata et al. (2002, 2003) have investigated equilibrium 
sequences of differentially rotating polytropes by using linear stability 
analysis, the numerical scheme of which is the same as that employed in this 
paper, and shown that the dynamical instability of a new type can set in even 
when $\beta$ is $\sim 0.05$ if the degree of differential rotation is very 
high.

However, even the classical $m = 2$ $f$-mode instability has not been fully 
studied by the linear stability analysis.  Thus, in this paper, we will study 
systematically  the dynamical stability of rapidly and differentially 
rotating polytropic stars by using the linear stability analysis.
It should be noted that the linear stability analysis of differentially 
rotating polytropes in this paper will give exact values of 
$\beta_{\rm critical}$ for the onset of $m = 2$ $f$-mode dynamical 
instabilities for the first time.

\section{Numerical Method}

The method to investigate stability by solving linearized basic equations
is the same as that used in Karino et al. (2000, 2001). Therefore, we will 
summarize the essence of the method briefly.

\subsection{Equilibrium states}\label{sec:2.1}

Axisymmetric equilibrium configurations of rotating Newtonian stars are 
obtained by solving the following equations in the spherical polar
coordinates $(r, \theta, \varphi)$:
\begin{equation}
\frac{1}{\rho}\nabla p
=-\nabla \phi + r \sin \theta \Omega ^2 \vec{e}_R \ ,
\end{equation}
\begin{equation}
\triangle \phi
=4 \pi G \rho  \ ,
\end{equation}
where $p$, $\rho$, $\phi$, $\Omega$, $\vec{e}_R$, and $G$ are the pressure, 
the density, the gravitational potential, the angular velocity, a unit vector 
in the $R$ direction of the cylindrical coordinates ($R, \varphi, z$), and 
the gravitational constant, respectively. As for the equation of state, we
assume the polytropic relation as follows:
\begin{equation}
p = K \rho^{1+\frac{1}{N}} \ ,
\end{equation}
where $K$ and $N$ are the polytropic constant and the polytropic index, 
respectively.  The rotation law is assumed as follows:
\begin{equation}
\Omega = \frac{\Omega_{\rm c} A^2}{ (r \sin\theta)^2 + A^2} \ , 
\label{eq:rot}
\end{equation}
where $\Omega_{\rm c}$ is the central angular velocity. The quantity $A$ is 
a parameter which represents the degree of differential rotation. The 
rotation becomes more differential as $A$ becomes smaller. We call this 
rotation law as the j-constant rotation law. 

One equilibrium state can be characterized by specifying three parameters as 
follows: the polytropic index $N$, the parameter $A$ and the 
parameter describing the amount of rotation which can be represented by the 
following axis ratio: 
\begin{equation}
q \equiv r_{\rm p} / r_{\rm e} \ ,
\end{equation}
where $r_{\rm p}$ and $r_{\rm e}$ are the polar and equatorial radii,
respectively. These equations are solved by applying the proper boundary 
conditions and using the SFNR method of Eriguchi and M\"{u}ller (1985).  
Equilibrium models in this paper have been solved by using mesh numbers 
$(r \times \theta) = (42 \times 19)$.

\subsection{Perturbed States}

As mentioned before, the linearized equations are solved by the scheme 
developed by  Karino et al. (2000, 2001). The perturbed quantities are 
expanded as follows:
\begin{equation}
\delta f (r,\theta,\varphi,t)
= \sum_m \exp(-i(\sigma t - m \varphi)) f_m(r,\theta),
\label{eq:expand}
\end{equation}
where $\delta f$ means the Euler perturbation of the corresponding quantity.
Here, $m$ is the azimuthal mode number and $\sigma$ is the eigenfrequency 
of the oscillation mode. As for the density and the pressure perturbations, 
we assume the adiabatic changes. Since we want to study the bar-type 
oscillations, we will consider only the oscillation modes with $m=2$ in this 
paper. 

In the stability analysis in this paper we have also used mesh numbers 
$(r \times \theta) = (42 \times 19)$. Although this mesh number seems to
be too small, the obtained eigenvalues can be considered to be accurate 
enough.  We have checked the eigenvalues by changing the mesh number.
For $N=1.0$ polytropes with extremely rapid rotation ($q = 0.2$), the 
accuracy of the obtained eigenfrequency can be estimated by 
$\delta \sigma(m,n) \equiv |(\sigma_m- \sigma_n) /
\sigma_m|$, where $\sigma_k$ denotes the eigenvalue when
the mesh number in the r-direction is $k$. For $A=1.0$ models, 
 $\delta \sigma (42,30)=$ 1.8 \%, and $\delta \sigma(42,36) = 0.6 $ \%.

\section{Numerical Results}

By fixing the values of $A$ and $N$ and by changing the value of $q$,
we can obtain eigenvalues of $m = 2$ $f$-mode oscillations along an 
equilibrium sequence. After that by changing the values of $A$ and/or $N$
and following the same procedure, we can obtain eigenstates for many
equilibrium sequences. In this paper, we have examined sequences with 
$N = 0.0, 1.0$ and 1.5, and mainly $ 1.0 \leq A^{-1}$. It may be noted that
for models with large values of $A$, mass begins to shed from the equatorial
surface before the dynamical instability sets in. Hence it is not necessary 
to consider larger values of $A$ more than the value for which mass shedding
occurs.

\begin{figure}
\plotone{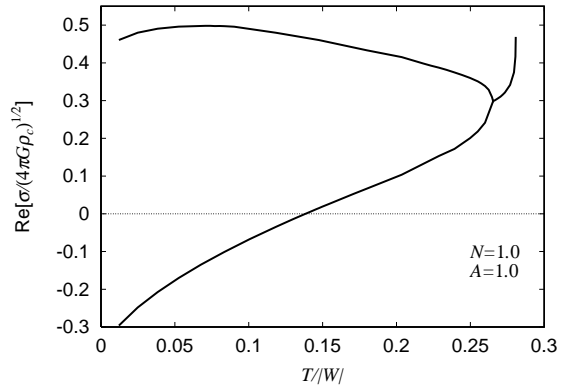}
\caption{Real part of the eigenvalue of the $m=2$ $f$-mode in units
of $\sqrt{4 \pi G \rho_{\rm c}}$ is plotted against the ratio $T/|W|$
for polytropes with $N=1.0$ and $A=1.0$. Here $\rho_c$ denotes the
central density. In slowly rotating regions,
the $f$-mode has two real branches, but when the stellar rotation exceeds
a certain critical point, these two branches merge into one branch,
in this real part of the eigenvalue plane, since eigenvalues are complex 
conjugate. This corresponds to the critical point where the $m = 2$ $f$-mode 
dynamical instability sets in.
\label{fig:1}}
\end{figure}

\begin{figure}
\plotone{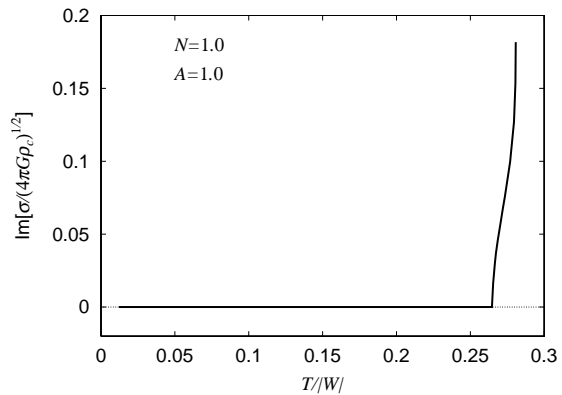}
\caption{Positive imaginary part of the eigenvalue
for the same mode as Figure \ref{fig:1} is plotted against $T/|W|$.
In slowly rotating regions, the imaginary part of the eigenfrequency vanishes
and it implies that the configurations are dynamically stable.
After the stellar rotation reaches a certain critical point, the imaginary 
part of the eigenvalues appears and grows very rapidly. This corresponds 
to the critical point where the  $m = 2$ $f$-mode dynamical instability sets in.
\label{fig:2}}
\end{figure}

In Figure \ref{fig:1}, the real part of eigenfrequencies of the
$N = 1.0$ sequences with $A = 1.0$ is plotted against the value of $\beta$.
In this plane, the $m=2$ $f$-mode has two branches for lower values of
$\beta$.  The upper branch corresponds to the co-rotating mode and the
lower one corresponds to the counter-rotating mode. These two branches merge 
into two complex conjugate branches for higher values of $\beta$. The 
positive imaginary part is shown in Figure \ref{fig:2}.
The model for which the two branches merge corresponds to the critical
configuration from which dynamically unstable configurations follow.
For this sequence,
i.e. $N = 1.0$ and $A=1.0$ sequence, the critical value where the dynamical
instability for $m = 2$ $f$-mode sets in is $\beta \simeq 0.266$.

\begin{figure}
\plotone{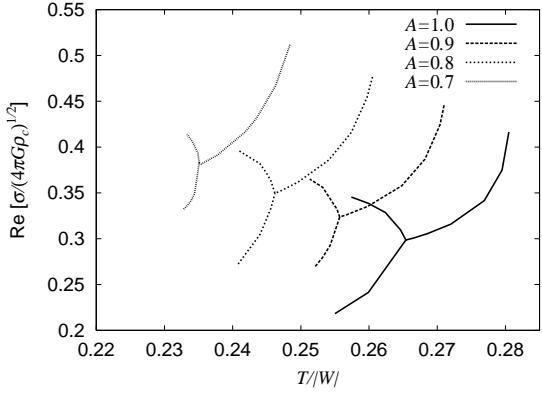}
\caption{Real part of the eigenvalues around the merging points is plotted
against $T/|W|$ for polytropic stars with $N = 1.0$ and $A = 0.7, 0.8, 0.9,
1.0$. As the degree of differential rotation becomes higher, 
the values of $T/|W|$ at the critical points become lower.
\label{fig:3}}
\end{figure}

\begin{figure}
\plotone{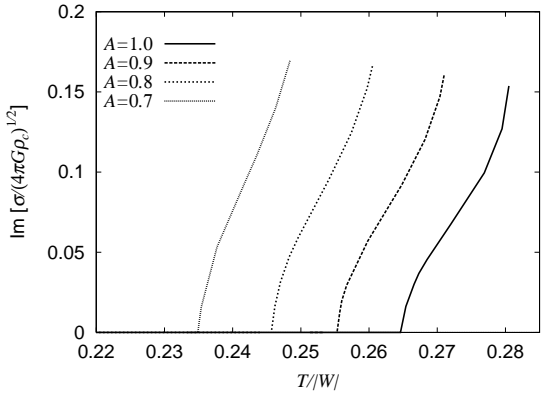}
\caption{Same as Figure \ref{fig:3} but for the imaginary part of eigenvalues.
\label{fig:4}}
\end{figure}

The eigenvalues of configurations of $N = 1$ polytropic sequences with 
several values of $A$ are shown against the value of $\beta$: real parts 
in Figure \ref{fig:3} and imaginary parts in Figure \ref{fig:4}.
It is clear from Figures \ref{fig:3} and \ref{fig:4} that the values of 
$\beta$ where two real modes change to two complex conjugate modes become 
smaller as the degree of differential rotation becomes higher. 
It means that the critical point of dynamical instability will appear for 
smaller values of $\beta$ as the degree of differential rotation becomes 
significant.

\begin{figure}
\plotone{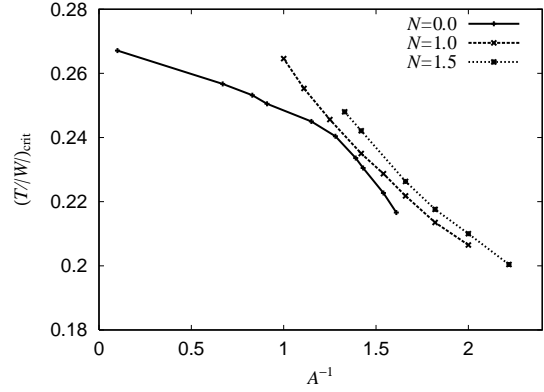}
\caption{Values of $T/|W|$ at the critical points where the $m = 2$ $f$-mode 
dynamical instability sets in are plotted against the degree of differential 
rotation, $A^{-1}$. The stellar models are polytropes with $N=0.0, 1.0$
and $1.5$. 
As the degree of differential rotation becomes higher, 
the critical value of $T/|W|$ tends to decrease.
\label{fig:5}}
\end{figure}

The dependency of the critical points of dynamical instability on the degree 
of differential rotation and the compressibility can clearly be seen from 
Figure \ref{fig:5}. In this figure, the values of $\beta$ at the critical 
points for $N = 0.0, 1.0$ and $1.5$ polytropes are plotted against the
parameter $A^{-1}$. The critical values of $\beta$ for the $m = 2$ $f$-mode 
dynamical instability depend significantly on the values of $A$ but depend only
slightly on the values of $N$ for the high degree of differential rotation.
Note that, as it should be,  the critical value of $\beta_{\rm critical}$
approaches to $\beta_{\rm critical} = 0.27$ when the rotation law tends to
the rigid rotation ($A^{-1} \to 0$) and the unperturbed configuration becomes 
the Maclaurin spheroid.

\begin{figure}
\plotone{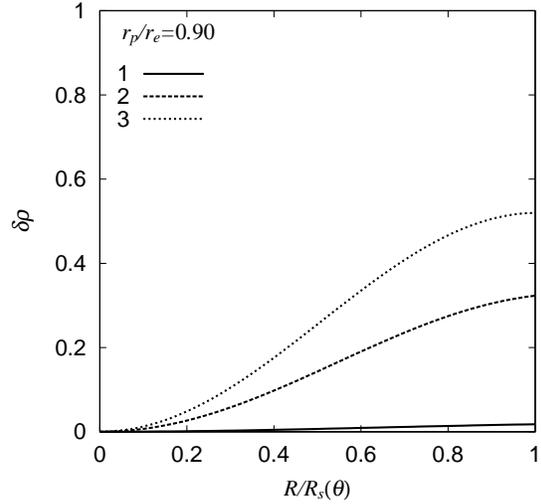}
\caption{Eigenfunctions of the density perturbation for the bar-mode 
oscillation of a differentially rotating polytropic star with $N = 1.0$, 
$A = 1.0$, and $q = 0.90$ are plotted against the normalized distance from 
the rotation axis. Here $R_{s}(\theta)$ is the radius to the surface along
the the polar angle $\theta$. Three curves correspond to the eigenfunctions 
on the spokes with $\theta = \pi / 6$ (curve 1), $\theta = \pi /3$ (curve 2), 
and $\theta = \pi/2$ (curve 3). The value of the eigenfunctions is
normalized so that the maximum absolute value among all the perturbed
quantities becomes unity.
\label{fig:6}}
\end{figure}

\begin{figure}
\plotone{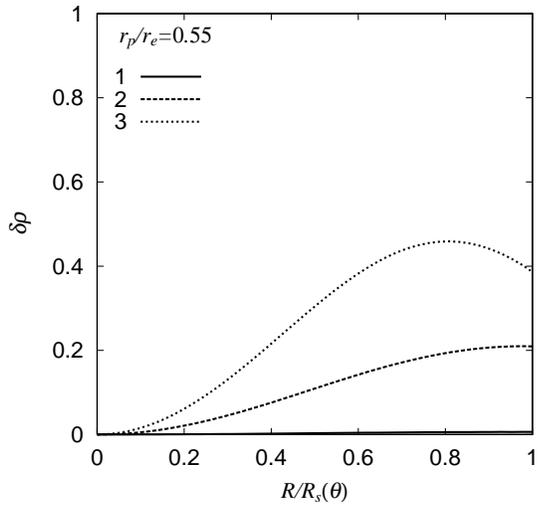}
\caption{Same as Figure \ref{fig:6} but for $q = 0.55$.
\label{fig:7}}
\end{figure}

\begin{figure}
\plotone{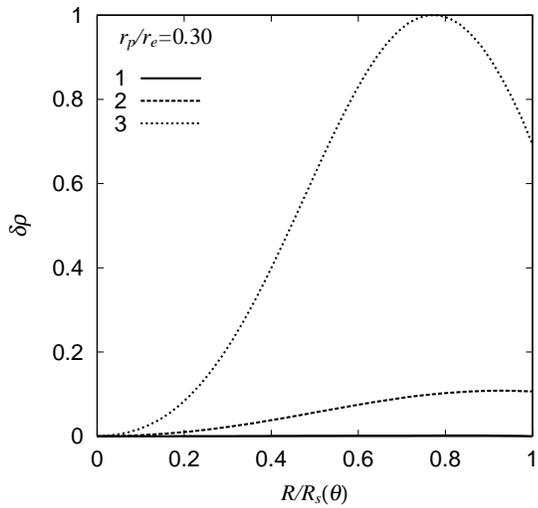}
\caption{Same as Figure \ref{fig:6} but for $q = 0.30$.
\label{fig:8}}
\end{figure}

In Figures \ref{fig:6}-\ref{fig:8}, the eigenfunctions of the density 
perturbation for $N = 1$ polytropes with $A = 1.0$ but with different 
values of $q$ are plotted against the normalized distance from the center
of the star at selected values of $\theta$.  In slowly rotating regions, 
the eigenfunctions are almost similar to those of $f$-mode oscillations
for the Maclaurin spheroid, i.e. monotonically increasing functions of 
the distance from the center (see e.g. Tassoul 1978).  As the stellar 
rotation becomes faster, however, the eigenfunctions on and near the 
equator attain their maximum values on the way to the surface but not
on the surface.

\section{Discussions}

By making use of the linear stability analysis, we have shown that, for 
$m = 2$ $f$-mode oscillations, higher degree of differential rotation makes 
the stars more unstable than previously thought.
Although results of recent dynamical simulations have shown the same tendency,
it should be noted that our present results have been obtained by
the systematic analysis of the oscillations and that we can understand the
occurrence of this type of instability from a more rigorous study.

Since the parameter regions for the existence of dynamically unstable 
configurations are widened, we need to reconsider some astrophysical 
situations under the light of the present results.
There are several situations where high differential rotation would be 
realized. One example can be outcomes of collapsing gases or stars,
i.e. star formation stages or core collapse stages of massive stars
or accretion-induced collapse processes of stars. 
Other example may be outcomes of collision or merging of two stars, i.e.
merging of two neutron stars in a binary neutron star system.

In the star formation process, if the angular momentum is not lost from the
gaseous system, the rotation of the final young stellar objects would be 
extremely rapid and highly differential. Thus they might suffer from 
the dynamical instability discussed in this paper. Although the outcomes
of the dynamical instability have not been fully understood yet, several 
authors have argued that dynamical instabilities during the star formation
stages are related to binary formations (see e.g. Bonnell 1994; 
Matsumoto \& Hanawa 1999), and others have suggested that they would
result in new systems consisting of a central young stellar object and a 
disk around it (see e.g. Bate 1998). In order to obtain a final answer to 
this problem, reliable nonlinear simulations and high-resolution 
infrared/radio observations about star forming regions are required. 

Another example is related to newly-born neutron stars. There are several
ways for neutron stars to be formed. One is the formation via core collapse
of massive stars followed by Type II supernova explosions. In recent 
simulations, it is shown that the compact object formed from an iron core 
collapse is rotating rapidly and differentially with $\beta \sim 0.2$ 
(Zwerger \& M\"{u}ller 1997; Dimmelmeier et al. 2002). Another is a merging 
of two neutron stars in a binary neutron star system. In the merging process 
in a binary neutron star system, there appears a rapidly rotating massive 
neutron star with high degree of differential rotation (see e.g. Shibata and 
Uryu 2000).  Neutron stars are also considered to be formed by accretion onto 
massive white dwarfs. In this accretion induced scenario, the outcome would be 
a rapidly and differentially rotating massive neutron star (see e.g. Liu
\& Lindblom 2000). Therefore, for these differentially rotating neutron
stars, it would be very likely that the dynamical instability would set in 
easily and lead the systems to nonaxisymmetric configurations just depending 
on the degree of differential rotation and irrespective of the compressibility. Those deformed and rapidly rotating compact objects would emit a large amount 
of gravitational waves. Therefore, such young neutron stars could be important 
targets of ground-based laser interferometric detectors of gravitational 
waves (see e.g. Brown 2000; Shibata et al. 2002)~\footnote{When we consider 
these compact stars, the effect of general relativity becomes important. 
In relation to the dynamical instability, however, it is only argued that the 
critical values might become slightly smaller by the general relativistic 
effect (Shibata, Baumgarte, \& Shapiro 2000).}.

Concerning the rotation law dependence of the critical values for the
$m = 2$ $f$-mode dynamical instability, we have performed linear stability
analysis for equilibrium models with the v-constant rotation law 
(see e.g. Karino et al. 2001). However, the effect of differential
rotation seems very weak and the dynamical instability occurs only 
for models with very stiff equations of state and/or extremely small values
of the parameter $A$. Additionally, the change of the critical values of 
$\beta$ is very little. The critical values for $N = 0.5$ polytropes 
are $T/|W| \sim 0.266$ for $A=0.8$ and $T/|W| \sim 0.260$ for $A=0.5$.

\section{Summary}

In this paper, we have performed linear stability analysis of equilibrium
sequences of differentially rotating polytropes in Newtonian gravity and 
obtained critical values of $\beta$, i.e. $\beta_{\rm critical}$, which 
determine the equilibrium states where the dynamical instability against 
$m = 2$ $f$-mode oscillations sets in. The rotation law which we have employed
is the j-constant law (Eq.~(\ref{eq:rot})). Our important finding is that the 
critical points of $m = 2$ $f$-mode dynamical instability depend strongly 
on the degree of differential rotation while they depend weakly on the 
compressibility. The critical values of $\beta_{\rm critical}$ decreases as 
the degree of differential rotation becomes higher. Therefore, we should not 
use the "universal" value of $\beta_{\rm critical} = 0.27$ for the stability
analysis of rapidly rotating objects with high degree of differential rotation 
from now on.

\acknowledgments

We are grateful to Drs. K. Uryu and S'i. Yoshida for their discussions and
useful advice. SK thanks Prof. Ohara and Prof. Nishi for their useful 
comments. This study is partially supported by the Research Fellowship 
of the Japan Society for the Promotion of Science for Young Scientists
and by the Grant-in-Aid for Scientific Research (C) of the JSPS (14540244).

\end{document}